\documentclass[structabstract]{aa}
\usepackage{graphicx}
\usepackage{longtable}
\usepackage[varg]{txfonts}
\usepackage[normalem]{ulem}

\begin{document}
    \title{Radiative data for highly excited 3d$^8$4d levels in Ni II from laboratory measurements and atomic calculations}

   \author{H. Hartman
           \inst{1,2},
           L. Engstr\"om
          \inst{3},
          H. Lundberg
          \inst{3},
          H. Nilsson
          \inst{2,3},
          P. Quinet
          \inst{4,5},
          V. Fivet
          \inst{4},
          P. Palmeri
          \inst{4},
          G. Malcheva
          \inst{6}
          \and
          K. Blagoev
          \inst{6}
%          C. Ptolemy\inst{2}\fnmsep\thanks{Just to show the usage
%          of the elements in the author field}
          }

   \institute{Material Sciences and Applied Mathematics, Malm\"o University, 20506 Malm\"o, Sweden
         \and
             Lund Observatory, Lund University, Box 43, SE-221 00 Lund, Sweden
         \and
             Department of Physics, Lund University, Box 118, SE-221 00 Lund, Sweden
         \and
             Physique Atomique et Astrophysique, Universit\'e de Mons, B-7000 Mons, Belgium
         \and
             IPNAS, Universit\'e de Li\`ege, B-4000 Li\`ege, Belgium
         \and
             Institute of Solid State Physics, Bulgarian Academy of Sciences, 72 Tzarigradsko Chaussee, BG-1784 Sofia, Bulgaria\\
             \email{Henrik.Hartman@mah.se}
%             \thanks{The university of heaven temporarily does not
%                    accept e-mails}
             }
\authorrunning{H. Hartman et al.}
\titlerunning{Radiative data for highly excited 3d$^8$4d levels in Ni II}
   \date{Received ??; accepted ??}
  \abstract
   {}
   {This work reports new experimental radiative lifetimes and calculated oscillator strengths for transitions from 3d$^8$4d levels of astrophysical interest in singly ionized nickel.}
   {Radiative lifetimes of seven high-lying levels of even parity in Ni II (98400 -- 100600 cm$^{-1}$) have been measured using the time-resolved laser-induced fluorescence method. Two-step photon excitation of ions produced by laser ablation has been utilized to populate the levels. Theoretical calculations of the radiative lifetimes of the measured levels and transition probabilities from these levels are reported. The calculations have been performed using a pseudo-relativistic Hartree-Fock method, taking into account core polarization effects.}
   {A new set of transition probabilities and oscillator strengths has been deduced for 477 Ni II transitions of astrophysical interest in the spectral range 194 -- 520 nm depopulating even parity 3d$^8$4d levels. 
   %They are supported by the good agreement between theory and experimental lifetime measurements.
   The new calculated gf-values are, on the average, about 20 \% higher than a previous calculation by Kurucz (http://kurucz.harvard.edu) and yield lifetimes within 5 \% of the experimental values.}
   {}

   \keywords{atomic data --
                methods: laboratory: atomic --
                techniques: spectroscopic
               }

   \maketitle
%
%________________________________________________________________

\section{Introduction}

The final stage of exothermal element production in massive stars is the iron-group elements, with nickel itself having the maximum binding energy per nucleon closely followed by iron. Higher $Z$ elements are produced by subsequent neutron capture. Nickel is therefore one of the most abundant iron-peak elements in cosmic objects. In addition, Nickel shows a line-rich spectrum due to its complex atomic structure and the lines appear in a variety of objects, from the interstellar medium and stars to the solar corona and supernova explosions. Nuclear statistical equilibrium models predict that iron and nickel are produced in high-temperature environments, i.e. explosive nucleosynthesis \citep{N03}, such as supernovae type Ia, where a white dwarf ignites, or supernovae type II, where the core of a massive star collapses into a neutron star \citep{SMS06}. The dominating product of these events is $^{56}$Ni, which is distributed to the surrounding gas during the outburst. Abundance determinations of nickel in stars serve as important constraints of stellar evolution and supernova explosion models. The current challenges for accurate elemental abundances are the development of 3D-model atmospheres and non-LTE modeling \citep{WJM11,LBA12}. Atomic data for levels of different excitation energies are important for this development. For example, in metal-rich stellar photospheres, transitions from low excitation states with a high population can be saturated whereas those from the less populated highly excited states are more likely to be optically thin. The present investigation of Ni II is part of an ongoing effort to provide such data for the second spectra of selected iron-group elements: Ti II \citep{LHE16}, Cr II \citep{ELN14}, Fe II \citep{HNE15} and Co II \citep{QFP16}.

There are two papers in the literature on experimental determination of radiative lifetimes in Ni II. \citet{LS87} used the Time-Resolved Laser-Induced Fluorescence (TR-LIF) method on a slow Ni$^+$ beam from a hollow-cathode source. Radiative lifetimes of 12 odd-parity levels of Ni II in the energy range from 51550 to 57080 cm$^{-1}$ are reported in that paper. Later, \citet{FL99} improved the experimental set-up, confirmed and extended the previous results to a total of 18 experimental lifetimes and reported transition probabilities for 59 lines in the VUV and UV spectrum of Ni$^+$, Ni II.

         \begin{figure*}[th]
   \centering
   \includegraphics[width=17cm]{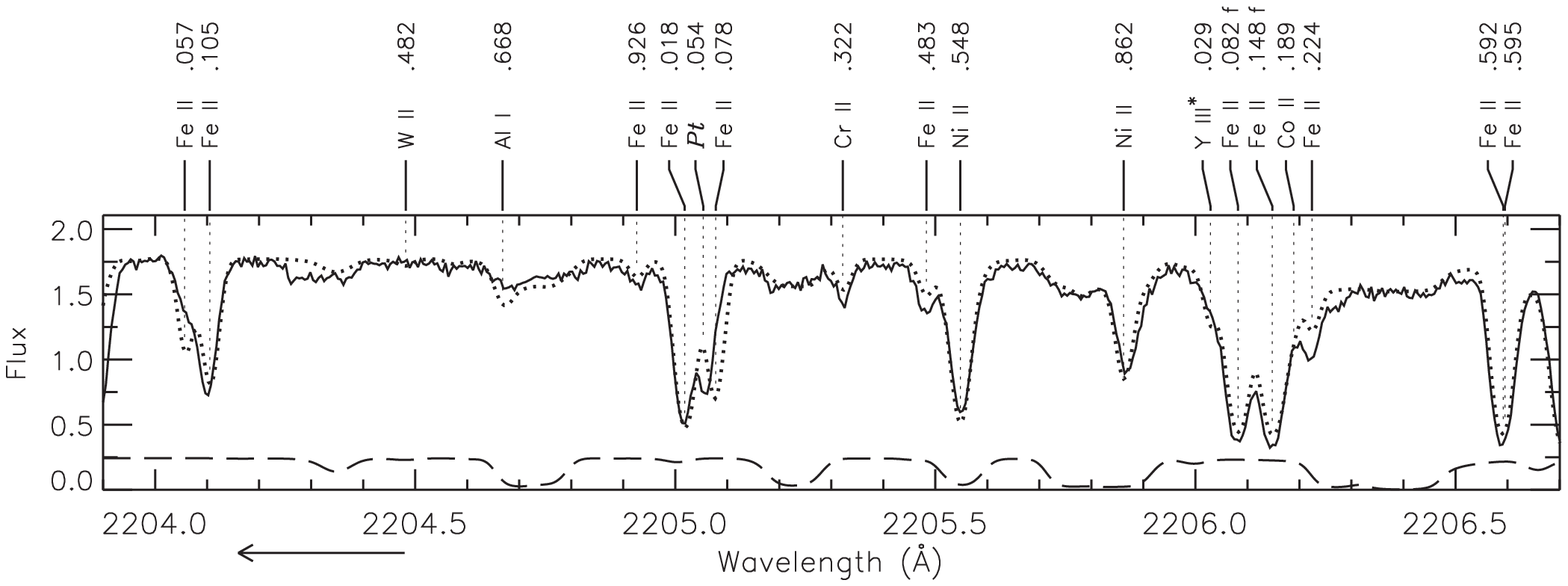}
      \caption{The $HST$/GHRS spectrum of Chi Lupi in the region around 2205 \AA, showing the prominent Ni II lines at 2205.548 and 2205.862 \AA\ studied in the present work. The solid line is the observed spectrum and the dashed line is the synthetic spectrum. Courtesy of \citet{BHB99} and reproduced by permission of the AAS. }
         \label{chilupi}
   \end{figure*}

In the present work, we have measured lifetimes for seven even-parity levels in the energy range 98400 to 100600 cm$^{-1}$. The lifetimes are measured using the TR-LIF method, and the high-lying states are populated using two-step photon excitation of ions produced by laser ablation. Furthermore, we report theoretical lifetimes obtained with a pseudo-relativistic Hartree-Fock method in good agreement with the experimental results, and calculated transition probabilities for 477 lines depopulating highly excited levels belonging to the even-parity 3d$^8$4d configuration in singly ionized nickel.

The lines studied in the present work appear strong in spectra of hot stars, such as the B9.5V type HgMn star Chi~Lupi~A \citep{BHB99}, where the lines form prominent absorption features despite their higher excitation. As an example, a spectral segment of Chi~Lupi from the $HST$/GHRS atlas is presented in Figure \ref{chilupi}. 
In addition, the Ni II lines are identified in several cooler template stars, e.g.\ the solar type star $\alpha$~Cen~A (spectral type G2~V) and Arcturus (K1.5~III) as reported by \citet{HWV05}. In the linelists of these stars, several Ni II lines are presented without multiplet number, indicating that they are of too high excitation to be listed in the multiplet table by \citet{M59}. All the Ni II lines in the 2000-3000 \AA\ region without multiplet number are included in our study. 

\section{Experiment}

The ground term in Ni II is [Ar] 3d$^9$ $^2$D, but as a starting point in the excitation schemes we used the $J$ = 9/2 and 7/2 levels in the second lowest term of even parity, 3d$^8$($^3$F)4s $^4$F, at 8394 and 9330 cm$^{-1}$, respectively \citep{S70}. These levels were populated directly in the plasma produced by the ablation laser. The first tuneable laser excited the intermediate, odd parity, levels in the 3d$^8$4p configuration around 55000 cm$^{-1}$, from where the final, even parity, levels in 3d$^8$4d around 100000 cm$^{-1}$ were reached with the second tuneable laser. The excitation and detection channels used in this work are given in Table 1.

The experimental set-up for two-step excitations at the Lund High Power Laser Facility has recently been described by \cite{LHE16}, and for an overview we refer to Figure 1 in that paper. Here we only give the most important details. Ni$^+$ ions in the 3d$^8$($^3$F)4s $^4$F term were created by focusing 532 nm, 10~ns long, laser pulses onto a rotating nickel target placed in a vacuum chamber with a pressure of about 10$^{-4}$ mbar. The two short wavelength excitation laser beams entered the vacuum chamber at a small relative angle and were focused on the expanding plasma plume about 5 mm above the target. All lasers operated at 10~Hz and were synchronized by a delay generator. The time resolved fluorescence from the excited states (both intermediate and final) was detected at right angles to the lasers by a 1/8~m grating monochromator, with its 0.28 mm wide entrance slit oriented parallel to the excitation laser beams and perpendicular to the ablation laser. The dispersed light was registered by a fast micro-channel-plate photomultiplier tube (Hamamatsu R3809U) and digitized by a Tektronix DPO 7254 oscilloscope. A second channel on the oscilloscope simultaneously registered the excitation pulse shape, as detected by a fast photo diode. The PM tube has a rise time of 200~ps and the oscilloscope sampled the decay and pulse shape at every 50~ps. All spectral measurements were performed in the second spectral order, which is closer to the optimum efficiency of the blazed grating, resulting in a linewidth of about 0.5~nm.

For both the first and the second excitations we used the frequency tripled output from Nd:YAG pumped dye lasers (Continuum Nd-60), primarily operating with DCM dye. The first step had a pulse length of 10~ns whereas, by injection seeding and compressing, for the second Nd:YAG laser a pulse length of about 1~ns could be obtained. Before every measurement the delay between the two lasers was checked and, if necessary, adjusted so that the short pulse from second laser coincided with the maximum population of the intermediate level. The short pulse length in the second step and the high time resolution of the detection system is necessary to accurately measure the short lifetimes involved (1.2 - 1.3~ns). To reach a sufficient statistical accuracy each decay curve was averaged over 1000 laser pulses. The final lifetime analysis used the code DECFIT \citep{PQF08} to fit a single exponential decay, convoluted by the measured pulse shape, and a background function to the observed decay curve. Typically 10 to 20 measurements, performed during different days, were averaged to obtain the final lifetimes, given in Table 2. The quoted uncertainties in the results are mainly due to the scatter between the individual measurements.

As discussed by \citet{LHE16}, two step measurements may lead to complicated blending situations that must be taken into account in the planning and execution of the experiment. In the Ni case, we note from Table 1 that the two excitation lasers as well as the fluorescence channels all occur in the narrow wavelength interval 210 - 226 nm. Thus, most of the recorded decay curves are influenced by the very intense decay from the intermediate levels. This contribution extends over more than 10ns, due to the length of the first step laser pulses, and is noticeable even at rather large wavelength differences. Fortunately, this effect may be accurately compensated for by subtracting a separate decay measurement, with the second step laser blocked, before the final lifetime analysis. A worse case is encountered in the measurement of the 4d $^4$H$_{13/2}$ level where also scattered light from the second laser influences the decay. This is illustrated in Figure 2, and corrected for by a 'background' measurement where the second laser is not blocked but detuned slightly from resonance. Finally, transitions from levels in the 3d$^8$4p configuration populated through the decay of the 4d level under investigation, so called cascades, may cause blending problems. Since this cannot be compensated for, one has to carefully choose the fluorescence channels to use, and if no sufficiently intense channels remain this particular level has to be omitted from the investigation. An example of this is the failure to measure the 4d $^4$G$_{7/2}$ level at 100475.8 cm$^{-1}$ since all strong decay channels are blended by cascades.

\section{Theoretical calculations}

Calculations of energy levels and radiative transition rates in Ni II have been carried out using the relativistic Hartree-Fock (HFR) approach \citep{C81} modified to take core-polarization effects into account \citep{QPB99,QPB02}. This method (HFR+CPOL) has been combined with a least-squares optimization process of the radial parameters to reduce the discrepancies between the Hamiltonian eigenvalues and the available experimental energy levels from \citet{S70}. The following 23 configurations were explicitly introduced in the calculations: 3d$^9$, 3d$^8$4d, 3d$^8$5d, 3d$^7$4s4d, 3d$^7$4s5d, 3d$^6$4s$^2$4d, 3d$^8$4s, 3d$^8$5s, 3d$^7$4s$^2$, 3d$^7$4s5s, 3d$^6$4s$^2$5s for the even parity and 3d$^8$4p, 3d$^8$5p, 3d$^7$4s4p, 3d$^7$4s5p, 3d$^6$4s$^2$4p, 3d$^6$4s$^2$5p, 3d$^8$4f, 3d$^8$5f, 3d$^7$4s4f, 3d$^7$4s5f, 3d$^6$4s$^2$4f, 3d$^6$4s$^2$5f for the odd parity.

The ionic core considered for the core-polarization model potential and the correction to the transition dipole operator was a 3d$^6$ Ni V core. The dipole polarizability, $\alpha$$_d$, for such a core is 0.94 a$_0^3$ according to \citet{FKS76}. We used the HFR mean radius of the outermost 3d core orbital, 1.004 a$_0$, for the cut-off radius.

For the 3d$^9$, 3d$^8$4d, 3d$^8$5d, 3d$^8$4s, 3d$^8$5s, 3d$^7$4s$^2$ even configurations and the 3d$^8$4p, 3d$^8$5p, 3d$^7$4s4p, 3d$^8$4f, 3d$^8$5f odd configurations, the average energies ($E$$_{av}$), the electrostatic direct ($F^k$) and exchange ($G^k$) integrals, the spin-orbit ($\zeta$$_{nl}$) and the effective interaction ($\alpha$) parameters were allowed to vary during the fitting process. All other Slater integrals were scaled down by a factor 0.80 following a well-established procedure \citep{C81}. The standard deviations of the fits were 212 cm$^{-1}$ for the even parity and 77 cm$^{-1}$ for the odd parity.

\section{Results and discussion}

The radiative lifetimes measured and computed in the present work are presented in Table 2. The theoretical lifetimes obtained in this work agree with the experimental values within about 5\%. Table 2 includes the theoretical lifetimes obtained by \citet{K11}. The latter work also used a semi-empirical approach based on a superposition of configurations calculation with a modified version of the Cowan codes \citep{C81} and experimental level energies \citep{S70} to improve the results. In this case the calculated values are about 11\% higher than the experiments.

Table 3 presents the computed oscillator strengths and transition probabilities for the strongest transitions (log $gf$ $>$ -4) depopulating the even 3d$^8$4d levels located in the range 98467--103664 cm$^{-1}$. This table also presents the cancellation factors ($CF$) as defined by \citet{C81}. Transitions with a $CF$ lower than 0.05 should be considered with great care as they are affected by cancellation effects.

Most of the previous experimental and theoretical studies of radiative parameters in Ni II were focused on the spectral lines from the odd parity 3d$^8$4p configuration \citep{ZF98,FL99,FDG00,FWL00,JT06,MAA11} and the 3d$^8$4s - 3d$^8$4p or 3d$^9$ - 3d$^8$4p transitions.  
Recently, an extensive calculation of atomic structure data for Ni II was published by \citet{CHR16}. This work resulted in transition rates and oscillator strengths for 5023 electric dipole lines involving the 3d$^9$, 3d$^8$4s, 3d$^7$4s$^2$, 3d$^8$4p and 3d$^7$4s4p configurations. 

To our knowledge, the only work listing also oscillator strengths from the 3d$^8$4d even levels is the database of \citet{K11}. 
Figure 3 shows a good general agreement between the two data sets. However, a closer inspection reveals that our new oscillator strengths are systematically higher than the values by \citet{K11}. 
The mean ratio $gf$(This work)/$gf$(Kurucz) is 1.22 for lines with log $gf$ $>$ -4. The new log $gf$ values are thus on average 20\% larger than the previous calculation by \citet{K11}.
The difference between the two sets of results is possibly due to different values of the radial dipole integrals in calculations of the line strengths. In the case of 3d$^8$4p -- 3d$^8$4d transition array for example, the reduced matrix element <4p||r$^1$||4d> computed in our work was -4.55664 a.u. and  -4.66532 a.u. with and without core-polarization, respectively, while, as far as we understand, Kurucz used a value scaled down to -4.47840 a.u. which tends to weaken the corresponding oscillator strengths.

\section{Conclusion}
We report seven new experimental radiative lifetimes for 3d$^8$4d levels in Ni II, measured by two-step excitation using time-resolved laser-induced fluorescence on a laser ablation plasma. In addition, we report an extensive theoretical study using a relativistic Hartree-Fock technique optimized on experimental level energies. The theoretical and experimental lifetimes agree within 5\%, which serves as benchmark for  the accuracy of the 477 calculated oscillator strengths for the strong transitions around 200-220 nm belonging to the 3d$^8$4p - 3d$^8$4d transition array. 
Furthermore, on a two-standard deviation level the new theoretical $gf$ values as well as the results from \citet{K11} agree with the experimental lifetimes.

\begin{acknowledgements}
This work has received funding from LASERLAB-EUROPE (grant agreement no. 284464, EC's Seventh Framework Programme), the Swedish Research Council through the Linnaeus grant to the Lund Laser Centre and a VR project grant 621-2011-4206 (H.H.), and the Knut and Alice Wallenberg Foundation. P.P. and P.Q. are respectively Research Associate and Research Director of the Belgian National Fund for Scientific Research F.R.S.-FNRS from which financial support is gratefully acknowledged. V.F. acknowledges the Belgian Scientific Policy (BELSPO) for her Return Grant. P.Q., V.F., P.P., G.M. and K.B. are grateful to the colleagues from Lund Laser Center for their kind hospitality and support.

\end{acknowledgements}

\bibliographystyle{aa} % style aa.bst
\bibliography{ni_ii_ref} % your references Yourfile.bib

\begin{thebibliography}{26}
\expandafter\ifx\csname natexlab\endcsname\relax\def\natexlab#1{#1}\fi

\bibitem[{{Brandt} {et~al.}(1999){Brandt}, {Heap}, {Beaver}, {Boggess},
  {Carpenter}, {Ebbets}, {Hutchings}, {Jura}, {Leckrone}, {Linsky}, {Maran},
  {Savage}, {Smith}, {Trafton}, {Walter}, {Weymann}, {Proffitt}, {Wahlgren},
  {Johansson}, {Nilsson}, {Brage}, {Snow}, \& {Ake}}]{BHB99}
{Brandt}, J.~C., {Heap}, S.~R., {Beaver}, E.~A., {et~al.} 1999, \aj, 117, 1505

\bibitem[{{Cassidy} {et~al.}(2016){Cassidy}, {Hibbert}, \&
  {Ramsbottom}}]{CHR16}
{Cassidy}, C.~M., {Hibbert}, A., \& {Ramsbottom}, C.~A. 2016, \aap, 587, A107

\bibitem[{{Cowan}(1981)}]{C81}
{Cowan}, R.~D. 1981, {The theory of atomic structure and spectra, } (University
  of California Press, Berkeley)

\bibitem[{{Engstr{\"o}m} {et~al.}(2014){Engstr{\"o}m}, {Lundberg}, {Nilsson},
  {Hartman}, \& {B{\"a}ckstr{\"o}m}}]{ELN14}
{Engstr{\"o}m}, L., {Lundberg}, H., {Nilsson}, H., {Hartman}, H., \&
  {B{\"a}ckstr{\"o}m}, E. 2014, \aap, 570, A34

\bibitem[{{Fedchak} \& {Lawler}(1999)}]{FL99}
{Fedchak}, J.~A. \& {Lawler}, J.~E. 1999, \apj, 523, 734

\bibitem[{{Fedchak} {et~al.}(2000){Fedchak}, {Wiese}, \& {Lawler}}]{FWL00}
{Fedchak}, J.~A., {Wiese}, L.~M., \& {Lawler}, J.~E. 2000, \apj, 538, 773

\bibitem[{Fraga {et~al.}(1976)Fraga, Karwowski, \& Saxena}]{FKS76}
Fraga, S., Karwowski, J., \& Saxena, K. 1976, Handbook of Atomic Data
  (Elsevier, Amsterdam)

\bibitem[{{Fritzsche} {et~al.}(2000){Fritzsche}, {Dong}, \& {Gaigalas}}]{FDG00}
{Fritzsche}, S., {Dong}, C.~Z., \& {Gaigalas}, G. 2000, Atomic Data and Nuclear
  Data Tables, 76, 155

\bibitem[{{Hartman} {et~al.}(2015){Hartman}, {Nilsson}, {Engstr{\"o}m}, \&
  {Lundberg}}]{HNE15}
{Hartman}, H., {Nilsson}, H., {Engstr{\"o}m}, L., \& {Lundberg}, H. 2015, \aap,
  584, A24

\bibitem[{{Hinkle} {et~al.}(2005){Hinkle}, {Wallace}, {Valenti}, \&
  {Ayres}}]{HWV05}
{Hinkle}, K., {Wallace}, L., {Valenti}, J., \& {Ayres}, T. 2005, {Ultraviolet
  Atlas of the Arcturus Spectrum, 1150-3800 A. } (San Francisco: ASP)

\bibitem[{{Jenkins} \& {Tripp}(2006)}]{JT06}
{Jenkins}, E.~B. \& {Tripp}, T.~M. 2006, \apj, 637, 548

\bibitem[{{Kurucz}(2011)}]{K11}
{Kurucz}, R. 2011, \url{http://kurucz.harvard.edu}

\bibitem[{{Lawler} \& {Salih}(1987)}]{LS87}
{Lawler}, J.~E. \& {Salih}, S. 1987, \pra, 35, 5046

\bibitem[{{Lind} {et~al.}(2012){Lind}, {Bergemann}, \& {Asplund}}]{LBA12}
{Lind}, K., {Bergemann}, M., \& {Asplund}, M. 2012, \mnras, 427, 50

\bibitem[{{Lundberg} {et~al.}(2016){Lundberg}, {Hartman}, {Engstr{\"o}m},
  {Nilsson}, {Persson}, {Palmeri}, {Quinet}, {Fivet}, {Malcheva}, \&
  {Blagoev}}]{LHE16}
{Lundberg}, H., {Hartman}, H., {Engstr{\"o}m}, L., {et~al.} 2016, \mnras

\bibitem[{{Manrique} {et~al.}(2011){Manrique}, {Aguilera}, \&
  {Arag{\'o}n}}]{MAA11}
{Manrique}, J., {Aguilera}, J.~A., \& {Arag{\'o}n}, C. 2011, \jqsrt, 112, 85

\bibitem[{{Moore}(1959)}]{M59}
{Moore}, C.~E. 1959, {A multiplet table of astrophysical interest. Part 1. }
  (NBS Technical Note, Washington: US Department of Commerce)

\bibitem[{{Nadyozhin}(2003)}]{N03}
{Nadyozhin}, D.~K. 2003, \mnras, 346, 97

\bibitem[{{Palmeri} {et~al.}(2008){Palmeri}, {Quinet}, {Fivet}, {Bi{\'e}mont},
  {Nilsson}, {Engstr{\"o}m}, \& {Lundberg}}]{PQF08}
{Palmeri}, P., {Quinet}, P., {Fivet}, V., {et~al.} 2008, \physscr, 78, 015304

\bibitem[{{Quinet} {et~al.}(2016){Quinet}, {Fivet}, {Palmeri}, {Engstr\"om},
  {Hartman}, {Lundberg}, \& {Nilsson}}]{QFP16}
{Quinet}, P., {Fivet}, V., {Palmeri}, P., {et~al.} 2016, \mnras, in press

\bibitem[{{Quinet} {et~al.}(2002){Quinet}, {Palmeri}, {Bi\'emont}, {Li},
  {Zhang}, \& {Svanberg}}]{QPB02}
{Quinet}, P., {Palmeri}, P., {Bi\'emont}, E., {et~al.} 2002, J. Alloys Comp.,
  344, 255

\bibitem[{{Quinet} {et~al.}(1999){Quinet}, {Palmeri}, {Bi{\'e}mont}, {McCurdy},
  {Rieger}, {Pinnington}, {Wickliffe}, \& {Lawler}}]{QPB99}
{Quinet}, P., {Palmeri}, P., {Bi{\'e}mont}, E., {et~al.} 1999, \mnras, 307, 934

\bibitem[{{Shenstone}(1970)}]{S70}
{Shenstone}, A. 1970, J.~Res.~Natl.~Bur.~Stand., 76A, 801

\bibitem[{{Stritzinger} {et~al.}(2006){Stritzinger}, {Mazzali}, {Sollerman}, \&
  {Benetti}}]{SMS06}
{Stritzinger}, M., {Mazzali}, P.~A., {Sollerman}, J., \& {Benetti}, S. 2006,
  \aap, 460, 793

\bibitem[{Wongwathanarat {et~al.}(2011)Wongwathanarat, Janka, \&
  M\"uller}]{WJM11}
Wongwathanarat, A., Janka, H.-T., \& M\"uller, E. 2011, in Proceedings of the
  International Astronomical Union, Vol.~7, Death of Massive Stars: Supernovae
  and Gamma-Ray Bursts, 150--153

\bibitem[{{Zsarg{\'o}} \& {Federman}(1998)}]{ZF98}
{Zsarg{\'o}}, J. \& {Federman}, S.~R. 1998, \apj, 498, 256

\end{thebibliography}
\clearpage
         \begin{figure*}[H]
   \centering
   \includegraphics[width=15cm]{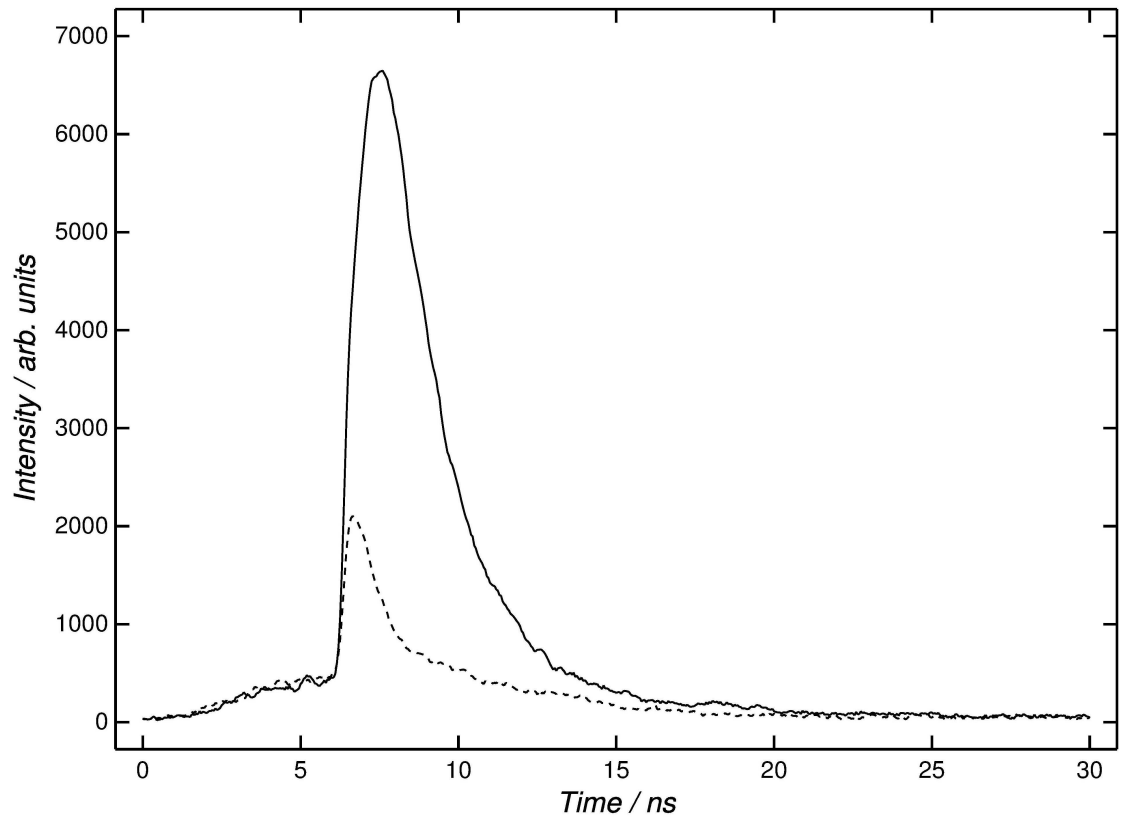}
      \caption{The solid line shows the first 30 ns of the decay of 4d $^4$H$_{13/2}$ in Ni II. The dashed line shows the combined background contribution from both the first- and second-step lasers, where the latter is detuned 0.04 nm from resonance. This background is substracted before the final lifetime analysis.}
         \label{Fig1}
   \end{figure*}
\clearpage
         \begin{figure*}[H]
   \centering
   \includegraphics[width=15cm]{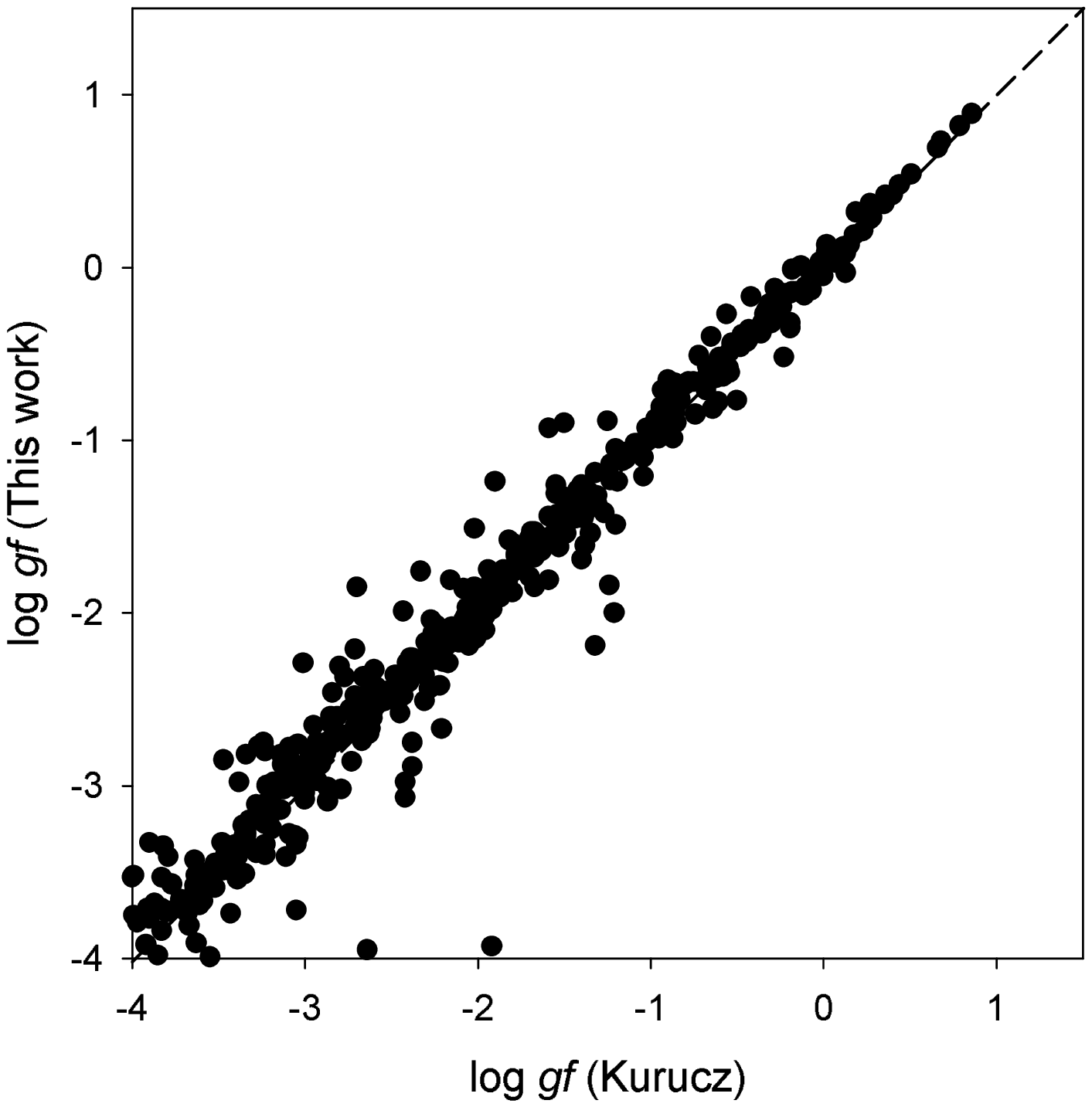}
      \caption{Comparison between the oscillator strengths (log $gf$) calculated in the present work and those reported by \citet{K11} for transitions from highly excited even-parity 3d$^8$4d levels in Ni II.}
         \label{Fig2}
   \end{figure*}
\clearpage
   \begin{table*}[H]
      \caption[]{Levels measured in the 3d$^8$($^3$F)4d configuration of Ni II and the corresponding excitation schemes.}
         \label{table:levels_exc}
         $$
         \begin{array}{cccccccccccc}
            \hline
            \noalign{\smallskip}
             & \multicolumn{3}{c}{$First~step~excitation$^a} & \multicolumn{2}{c}{$Second~step~excitation$} &  & \multicolumn{2}{c}{$Detection$} \\
             $Final level$ & $Start level$ & $Intermediate$ & \lambda_{air} & $Final level$ & \lambda_{air} &  $Scheme$^c & $Lower level$ & \lambda_{air} \\
                          & E$$^\mathrm{b} $(cm$^{-1}$)$ & E$$^\mathrm{b} $(cm$^{-1}$)$ & $(nm)$ & E$$^\mathrm{b}$ (cm$^{-1}$)$ & $(nm)$    & & E$$^\mathrm{b} $(cm$^{-1}$)$ & $(nm)$ \\
            \noalign{\smallskip}
            \hline
            \noalign{\smallskip} 
            ^4$D$_{7/2} & 8393.9 & 54557.0 & 216.6 & 98467.25 & 227.7 & 3\omega+S & 51557.8 & 213^d \\
            ^4$H$_{13/2}& 8393.9 & 53496.5 & 221.6 & 98822.55 & 220.6 & 3\omega   & 53496.5 & 221^e \\
            ^4$G$_{11/2}& 8393.9 & 53496.5 & 221.6 & 99132.78 & 219.0 & 3\omega   & 54557.0 & 224^d \\
            ^4$F$_{9/2} & 8393.9 & 53496.5 & 221.6 & 99154.81 & 218.9 & 3\omega   & 51557.8 & 210, 224^d \\
                        &        & 54557.0 & 216.6 &          & 224.1 & 3\omega+S & 54557.0 & 210 \\
            ^4$D$_{5/2} & 9330.0 & 55417.8 & 216.9 & 99559.33 & 226.5 & 3\omega   & 52738.4 & 214^d \\
            ^4$H$_{11/2}& 9330.0 & 55299.6 & 217.5 &100309.29 & 222.1 & 3\omega   & 55299.6 & 222^d \\
            ^4$G$_{9/2} & 8393.9 & 54557.0 & 216.6 &100619.26 & 217.0 & 3\omega   & 56371.4 & 226 \\
             \noalign{\smallskip}
            \hline
         \end{array}
         $$
\begin{list}{}{}
\item[$^{\mathrm{a}}$] For all measured levels, the first excitation step used the frequency tripled (3$\omega$) output from the dye laser
\item[$^{\mathrm{b}}$] \citet{S70} 
\item[$^{\mathrm{c}}$] S imply one added Stokes component of 4153 cm$^{-1}$
\item[$^{\mathrm{d}}$] Corrected for fluorescence background from the intermediate level
\item[$^{\mathrm{e}}$] Corrected for scattered light from both lasers, see also Figure 2
\end{list}
   \end{table*}
\clearpage
  \begin{table*}[H]
      \caption[]{Radiative lifetimes (in ns) for selected energy levels belonging to the 3d$^8$($^3$F)4d configuration of Ni II.}
         \label{table:levels_results}
         $$
         \begin{array}{cccccccccc}
            \hline
            \noalign{\smallskip}
            $Level$     & $Energy$^\mathrm{a}    & $Experiment$ & \multicolumn{2}{c}{$Calculations$} \\
                        & $(cm$^{-1}$)$ & $This work$ & $This work$ & $Kurucz$^\mathrm{b} \\
            \noalign{\smallskip}
            \hline
            \noalign{\smallskip}
            ^4$D$_{7/2}  &  98467.25 & 1.28 \pm 0.1 & 1.30 & 1.43 \\
            ^4$H$_{13/2} &  98822.55 & 1.25 \pm 0.1 & 1.32 & 1.41 \\
            ^4$G$_{11/2} &  99132.78 & 1.32 \pm 0.1 & 1.34 & 1.43 \\
            ^4$F$_{9/2}  &  99154.81 & 1.20 \pm 0.1 & 1.29 & 1.39 \\
            ^4$D$_{5/2}  &  99559.33 & 1.37 \pm 0.1 & 1.30 & 1.42 \\
            ^4$H$_{11/2} & 100309.29 & 1.30 \pm 0.1 & 1.33 & 1.41 \\
            ^4$G$_{9/2}  & 100619.26 & 1.25 \pm 0.1 & 1.35 & 1.44 \\
             \noalign{\smallskip}
            \hline
         \end{array}
         $$
\begin{list}{}{}
\item[$^{\mathrm{a}}$]  \citet{S70}
\item[$^{\mathrm{b}}$]  \citet{K11}
\end{list}
   \end{table*}
\onecolumn
\begin{longtable}{rrcrcrrr}
\caption{\label{table:values} Transition probabilities and oscillator strengths for spectral lines depopulating highly excited levels belonging to the even-parity 3d$^8$4d configuration of Ni II. xE+y stands for x $\times$ 10$^y$. Only transitions with log $gf$ $\ge$ -4.0 are listed in the table.}\\
\hline\hline
$\lambda$$^a$ (nm)  & \multicolumn{2}{c}{Lower odd level$^b$}        & \multicolumn{2}{c}{Upper 4d level$^b$} & \multicolumn{3}{c}{HFR+CPOL$^c$}  \\
                    & $E$ (cm$^{-1}$) & $J$                 & $E$ (cm$^{-1}$) & $J$          & log $gf$ & $gA$ (s$^{-1}$) & $CF$ \\
\hline
\endfirsthead
\caption{continued.}\\
\hline\hline
%$\lambda$  & \multicolumn{2}{c}{Lower level} & \multicolumn{2}{c}{Upper level} & \multicolumn{2}{c}{HFR+CPOL} & %\multicolumn{2}{c}{Normalized} \\
$\lambda$$^a$ (nm)  & \multicolumn{2}{c}{Lower odd level$^b$}        & \multicolumn{2}{c}{Upper 4d level$^b$} & \multicolumn{3}{c}{HFR+CPOL$^c$}  \\
                    & $E$ (cm$^{-1}$) & $J$                 & $E$ (cm$^{-1}$) & $J$          & log $gf$ & $gA$ (s$^{-1}$) & $CF$ \\
\hline
\endhead
\hline
\endfoot
194.2965	&	51558	&	3.5	&	103026	&	2.5	&	-2.69	&	3.64E+06	&	0.018	\\	
196.3670	&	52738	&	2.5	&	103664	&	1.5	&	-3.74	&	3.12E+05	&	0.003	\\	
198.8580	&	52738	&	2.5	&	103026	&	2.5	&	-1.82	&	2.58E+07	&	0.077	\\	
199.2730	&	51558	&	3.5	&	101740	&	3.5	&	-3.04	&	1.54E+06	&	0.004	\\	
199.8845	&	53635	&	1.5	&	103664	&	1.5	&	-2.23	&	9.76E+06	&	0.100	\\	
200.7049	&	51558	&	3.5	&	101366	&	2.5	&	-3.20	&	1.04E+06	&	0.003	\\	
200.7409	&	51558	&	3.5	&	101357	&	4.5	&	-2.21	&	1.02E+07	&	0.054	\\	
201.1847	&	51558	&	3.5	&	101247	&	2.5	&	-3.36	&	7.19E+05	&	0.015	\\	
201.6016	&	51558	&	3.5	&	101145	&	3.5	&	-2.86	&	2.25E+06	&	0.009	\\	
202.0071	&	54176	&	0.5	&	103664	&	1.5	&	-2.96	&	1.80E+06	&	0.057	\\	
202.4010	&	53635	&	1.5	&	103026	&	2.5	&	-1.97	&	1.75E+07	&	0.237	\\	
203.7607	&	51558	&	3.5	&	100619	&	4.5	&	-0.93	&	1.90E+08	&	0.160	\\	
203.8699	&	51558	&	3.5	&	100593	&	3.5	&	-2.67	&	3.44E+06	&	0.002	\\	
204.0085	&	52738	&	2.5	&	101740	&	3.5	&	-1.19	&	1.03E+08	&	0.166	\\	
204.3583	&	51558	&	3.5	&	100476	&	3.5	&	-2.36	&	7.01E+06	&	0.020	\\	
204.7195	&	51558	&	3.5	&	100390	&	2.5	&	-1.91	&	1.96E+07	&	0.022	\\	
204.9606	&	51558	&	3.5	&	100332	&	4.5	&	-3.71	&	3.10E+05	&	0.004	\\	
205.5061	&	55019	&	2.5	&	103664	&	1.5	&	-2.47	&	5.32E+06	&	0.222	\\	
205.5783	&	52738	&	2.5	&	101366	&	2.5	&	-1.57	&	4.25E+07	&	0.031	\\	
206.0365	&	52738	&	2.5	&	101258	&	1.5	&	-2.17	&	1.06E+07	&	0.016	\\	
206.0817	&	52738	&	2.5	&	101247	&	2.5	&	-1.82	&	2.40E+07	&	0.082	\\	
206.5192	&	52738	&	2.5	&	101145	&	3.5	&	-1.53	&	4.61E+07	&	0.090	\\	
206.6519	&	53365	&	4.5	&	101740	&	3.5	&	-3.00	&	1.54E+06	&	0.003	\\	
208.2372	&	55019	&	2.5	&	103026	&	2.5	&	-3.28	&	8.07E+05	&	0.021	\\	
208.2606	&	51558	&	3.5	&	99559	&	2.5	&	-0.93	&	1.79E+08	&	0.041	\\	
208.3016	&	53365	&	4.5	&	101357	&	4.5	&	-3.84	&	2.20E+05	&	0.000	\\	
208.7672	&	51558	&	3.5	&	99443	&	4.5	&	-0.40	&	6.10E+08	&	0.347	\\	
208.9002	&	52738	&	2.5	&	100593	&	3.5	&	-0.25	&	8.54E+08	&	0.424	\\	
209.2143	&	51558	&	3.5	&	99341	&	3.5	&	-1.07	&	1.31E+08	&	0.252	\\	
209.2284	&	53365	&	4.5	&	101145	&	3.5	&	-3.53	&	4.45E+05	&	0.001	\\	
209.3466	&	52738	&	2.5	&	100491	&	1.5	&	-0.58	&	3.99E+08	&	0.116	\\	
209.4129	&	52738	&	2.5	&	100476	&	3.5	&	-0.90	&	1.89E+08	&	0.453	\\	
209.4386	&	53635	&	1.5	&	101366	&	2.5	&	-0.44	&	5.46E+08	&	0.354	\\	
209.7923	&	52738	&	2.5	&	100390	&	2.5	&	-2.29	&	7.72E+06	&	0.004	\\	
209.9142	&	53635	&	1.5	&	101258	&	1.5	&	-1.40	&	6.04E+07	&	0.045	\\	
209.9611	&	53635	&	1.5	&	101247	&	2.5	&	-1.02	&	1.43E+08	&	0.460	\\	
209.9665	&	53635	&	1.5	&	101246	&	0.5	&	-0.65	&	3.38E+08	&	0.605	\\	
209.9832	&	55418	&	3.5	&	103026	&	2.5	&	-1.45	&	5.42E+07	&	0.098	\\	
210.0308	&	51558	&	3.5	&	99155	&	4.5	&	-0.05	&	1.35E+09	&	0.675	\\	
210.0693	&	56075	&	2.5	&	103664	&	1.5	&	-2.28	&	7.95E+06	&	0.036	\\	
210.5587	&	54263	&	3.5	&	101740	&	3.5	&	-2.98	&	1.56E+06	&	0.001	\\	
211.1695	&	52738	&	2.5	&	100079	&	1.5	&	-0.17	&	1.01E+09	&	0.619	\\	
211.5549	&	53365	&	4.5	&	100619	&	4.5	&	-1.97	&	1.59E+07	&	0.004	\\	
211.6225	&	56424	&	1.5	&	103664	&	1.5	&	-1.52	&	4.55E+07	&	0.287	\\	
211.6727	&	53365	&	4.5	&	100593	&	3.5	&	-1.53	&	4.44E+07	&	0.045	\\	
211.7490	&	53635	&	1.5	&	100845	&	0.5	&	-1.84	&	2.17E+07	&	0.015	\\	
211.8727	&	54557	&	4.5	&	101740	&	3.5	&	-3.53	&	4.39E+05	&	0.001	\\	
212.1445	&	53496	&	5.5	&	100619	&	4.5	&	-2.80	&	2.33E+06	&	0.005	\\	
212.1991	&	53365	&	4.5	&	100476	&	3.5	&	-1.91	&	1.83E+07	&	0.030	\\	
212.2313	&	54263	&	3.5	&	101366	&	2.5	&	-1.60	&	3.74E+07	&	0.029	\\	
212.2716	&	54263	&	3.5	&	101357	&	4.5	&	-2.67	&	3.14E+06	&	0.001	\\	
212.3294	&	54176	&	0.5	&	101258	&	1.5	&	-0.14	&	1.06E+09	&	0.836	\\	
212.3829	&	54176	&	0.5	&	101246	&	0.5	&	-0.93	&	1.72E+08	&	0.381	\\	
212.6835	&	51558	&	3.5	&	98561	&	2.5	&	0.48	&	4.47E+09	&	0.859	\\	
212.7678	&	54263	&	3.5	&	101247	&	2.5	&	-1.36	&	6.45E+07	&	0.108	\\	
212.8486	&	53365	&	4.5	&	100332	&	4.5	&	-1.71	&	2.83E+07	&	0.013	\\	
212.9239	&	56075	&	2.5	&	103026	&	2.5	&	-1.90	&	1.85E+07	&	0.073	\\	
212.9520	&	53365	&	4.5	&	100309	&	5.5	&	-1.21	&	9.12E+07	&	0.009	\\	
213.1096	&	51558	&	3.5	&	98467	&	3.5	&	0.41	&	3.78E+09	&	0.736	\\	
213.2342	&	54263	&	3.5	&	101145	&	3.5	&	-1.29	&	7.50E+07	&	0.024	\\	
213.3510	&	53635	&	1.5	&	100491	&	1.5	&	0.09	&	1.82E+09	&	0.775	\\	
213.5126	&	52738	&	2.5	&	99559	&	2.5	&	0.30	&	2.90E+09	&	0.744	\\	
213.5494	&	53496	&	5.5	&	100309	&	5.5	&	-1.86	&	2.02E+07	&	0.013	\\	
213.8139	&	53635	&	1.5	&	100390	&	2.5	&	-0.23	&	8.53E+08	&	0.708	\\	
213.9665	&	55019	&	2.5	&	101740	&	3.5	&	-1.12	&	1.10E+08	&	0.182	\\	
214.2068	&	54176	&	0.5	&	100845	&	0.5	&	-0.14	&	1.07E+09	&	0.852	\\	
214.2756	&	56371	&	3.5	&	103026	&	2.5	&	-1.64	&	3.34E+07	&	0.075	\\	
214.5150	&	52738	&	2.5	&	99341	&	3.5	&	-0.38	&	6.08E+08	&	0.617	\\	
214.5197	&	56424	&	1.5	&	103026	&	2.5	&	-2.15	&	1.02E+07	&	0.211	\\	
214.5819	&	54557	&	4.5	&	101145	&	3.5	&	-2.94	&	1.67E+06	&	0.011	\\	
215.2446	&	53635	&	1.5	&	100079	&	1.5	&	-0.78	&	2.40E+08	&	0.187	\\	
215.2610	&	55300	&	4.5	&	101740	&	3.5	&	-3.29	&	7.29E+05	&	0.001	\\	
215.5631	&	53635	&	1.5	&	100010	&	0.5	&	-0.14	&	1.04E+09	&	0.866	\\	
215.6511	&	54263	&	3.5	&	100619	&	4.5	&	-0.99	&	1.46E+08	&	0.038	\\	
215.6939	&	55019	&	2.5	&	101366	&	2.5	&	-0.61	&	3.51E+08	&	0.149	\\	
215.7734	&	54263	&	3.5	&	100593	&	3.5	&	-2.00	&	1.44E+07	&	0.007	\\	
215.8103	&	55418	&	3.5	&	101740	&	3.5	&	-1.81	&	2.21E+07	&	0.009	\\	
215.8464	&	54176	&	0.5	&	100491	&	1.5	&	-0.88	&	1.89E+08	&	0.238	\\	
215.9041	&	52738	&	2.5	&	99041	&	1.5	&	0.03	&	1.54E+09	&	0.868	\\	
216.1795	&	57420	&	2.5	&	103664	&	1.5	&	-0.49	&	4.60E+08	&	0.318	\\	
216.1983	&	55019	&	2.5	&	101258	&	1.5	&	-0.63	&	3.32E+08	&	0.435	\\	
216.2481	&	55019	&	2.5	&	101247	&	2.5	&	-0.18	&	9.38E+08	&	0.674	\\	
216.3205	&	54263	&	3.5	&	100476	&	3.5	&	0.10	&	1.78E+09	&	0.619	\\	
216.7253	&	54263	&	3.5	&	100390	&	2.5	&	-0.64	&	3.26E+08	&	0.291	\\	
216.7298	&	55019	&	2.5	&	101145	&	3.5	&	0.21	&	2.26E+09	&	0.369	\\	
216.9567	&	53365	&	4.5	&	99443	&	4.5	&	-0.52	&	4.32E+08	&	0.139	\\	
216.9955	&	54263	&	3.5	&	100332	&	4.5	&	0.41	&	3.60E+09	&	0.495	\\	
217.0296	&	54557	&	4.5	&	100619	&	4.5	&	-3.95	&	1.58E+05	&	0.000	\\	
217.0516	&	55300	&	4.5	&	101357	&	4.5	&	-1.45	&	5.07E+07	&	0.030	\\	
217.1535	&	54557	&	4.5	&	100593	&	3.5	&	-1.57	&	3.83E+07	&	0.043	\\	
217.4396	&	53365	&	4.5	&	99341	&	3.5	&	-0.32	&	6.76E+08	&	0.732	\\	
217.5677	&	55418	&	3.5	&	101366	&	2.5	&	-2.36	&	6.21E+06	&	0.005	\\	
217.5769	&	53496	&	5.5	&	99443	&	4.5	&	-0.89	&	1.81E+08	&	0.397	\\	
217.5832	&	57081	&	3.5	&	103026	&	2.5	&	-0.15	&	9.98E+08	&	0.540	\\	
217.6100	&	55418	&	3.5	&	101357	&	4.5	&	-0.87	&	1.92E+08	&	0.055	\\	
217.6795	&	53635	&	1.5	&	99559	&	2.5	&	-1.07	&	1.20E+08	&	0.076	\\	
217.7077	&	54557	&	4.5	&	100476	&	3.5	&	-2.10	&	1.13E+07	&	0.035	\\	
217.7847	&	54176	&	0.5	&	100079	&	1.5	&	-1.37	&	5.97E+07	&	0.095	\\	
218.1108	&	54176	&	0.5	&	100010	&	0.5	&	-1.54	&	4.02E+07	&	0.042	\\	
218.1316	&	55418	&	3.5	&	101247	&	2.5	&	-2.39	&	5.70E+06	&	0.010	\\	
218.1638	&	52738	&	2.5	&	98561	&	2.5	&	-1.62	&	3.36E+07	&	0.011	\\	
218.3217	&	53365	&	4.5	&	99155	&	4.5	&	0.37	&	3.25E+09	&	0.754	\\	
218.3913	&	54557	&	4.5	&	100332	&	4.5	&	-1.76	&	2.43E+07	&	0.045	\\	
218.4268	&	53365	&	4.5	&	99133	&	5.5	&	-0.35	&	6.22E+08	&	0.117	\\	
218.5002	&	54557	&	4.5	&	100309	&	5.5	&	-3.01	&	1.36E+06	&	0.000	\\	
218.6122	&	52738	&	2.5	&	98467	&	3.5	&	-1.02	&	1.34E+08	&	0.090	\\	
218.6217	&	55418	&	3.5	&	101145	&	3.5	&	-1.26	&	7.63E+07	&	0.035	\\	
218.9176	&	56075	&	2.5	&	101740	&	3.5	&	-0.69	&	2.83E+08	&	0.136	\\	
218.9497	&	53496	&	5.5	&	99155	&	4.5	&	-0.32	&	6.61E+08	&	0.876	\\	
219.0554	&	53496	&	5.5	&	99133	&	5.5	&	0.13	&	1.86E+09	&	0.732	\\	
219.2037	&	57420	&	2.5	&	103026	&	2.5	&	-0.21	&	8.65E+08	&	0.205	\\	
219.2092	&	53365	&	4.5	&	98969	&	5.5	&	0.73	&	7.40E+09	&	0.765	\\	
219.3535	&	55019	&	2.5	&	100593	&	3.5	&	-0.01	&	1.35E+09	&	0.770	\\	
219.8423	&	53496	&	5.5	&	98969	&	5.5	&	0.13	&	1.87E+09	&	0.918	\\	
219.8457	&	55019	&	2.5	&	100491	&	1.5	&	-1.05	&	1.23E+08	&	0.574	\\	
219.9189	&	55019	&	2.5	&	100476	&	3.5	&	0.01	&	1.40E+09	&	0.397	\\	
220.1659	&	53635	&	1.5	&	99041	&	1.5	&	-3.92	&	1.66E+05	&	0.000	\\	
220.3373	&	55019	&	2.5	&	100390	&	2.5	&	-0.24	&	7.93E+08	&	0.559	\\	
220.3467	&	56371	&	3.5	&	101740	&	3.5	&	0.06	&	1.58E+09	&	0.551	\\	
220.5548	&	53496	&	5.5	&	98823	&	6.5	&	0.89	&	1.06E+10	&	0.928	\\	
220.5862	&	55300	&	4.5	&	100619	&	4.5	&	0.28	&	2.62E+09	&	0.710	\\	
220.6978	&	54263	&	3.5	&	99559	&	2.5	&	-0.36	&	6.06E+08	&	0.702	\\	
220.7142	&	55300	&	4.5	&	100593	&	3.5	&	-0.81	&	2.12E+08	&	0.283	\\	
220.7262	&	56075	&	2.5	&	101366	&	2.5	&	-0.21	&	8.43E+08	&	0.224	\\	
221.1630	&	55418	&	3.5	&	100619	&	4.5	&	-0.01	&	1.33E+09	&	0.245	\\	
221.2544	&	56075	&	2.5	&	101258	&	1.5	&	-2.19	&	8.86E+06	&	0.008	\\	
221.2668	&	54263	&	3.5	&	99443	&	4.5	&	0.08	&	1.65E+09	&	0.636	\\	
221.2867	&	55300	&	4.5	&	100476	&	3.5	&	-0.64	&	3.12E+08	&	0.587	\\	
221.2917	&	55418	&	3.5	&	100593	&	3.5	&	0.03	&	1.45E+09	&	0.423	\\	
221.3066	&	56075	&	2.5	&	101247	&	2.5	&	-1.32	&	6.60E+07	&	0.035	\\	
221.3155	&	58493	&	2.5	&	103664	&	1.5	&	-0.41	&	5.31E+08	&	0.319	\\	
221.6502	&	53365	&	4.5	&	98467	&	3.5	&	-0.46	&	4.68E+08	&	0.732	\\	
221.7690	&	54263	&	3.5	&	99341	&	3.5	&	-0.04	&	1.25E+09	&	0.589	\\	
221.8111	&	56075	&	2.5	&	101145	&	3.5	&	0.42	&	3.57E+09	&	0.943	\\	
221.8569	&	55019	&	2.5	&	100079	&	1.5	&	-1.11	&	1.04E+08	&	0.456	\\	
221.8671	&	55418	&	3.5	&	100476	&	3.5	&	-1.51	&	4.19E+07	&	0.016	\\	
221.9930	&	55300	&	4.5	&	100332	&	4.5	&	-0.71	&	2.66E+08	&	0.141	\\	
222.1055	&	55300	&	4.5	&	100309	&	5.5	&	0.82	&	8.89E+09	&	0.955	\\	
222.1791	&	56371	&	3.5	&	101366	&	2.5	&	-1.30	&	6.83E+07	&	0.257	\\	
222.2233	&	56371	&	3.5	&	101357	&	4.5	&	0.54	&	4.62E+09	&	0.640	\\	
222.2929	&	55418	&	3.5	&	100390	&	2.5	&	-1.58	&	3.61E+07	&	0.026	\\	
222.3629	&	58706	&	1.5	&	103664	&	1.5	&	0.09	&	1.67E+09	&	0.679	\\	
222.4415	&	56424	&	1.5	&	101366	&	2.5	&	0.19	&	2.11E+09	&	0.859	\\	
222.5161	&	53635	&	1.5	&	98561	&	2.5	&	-1.94	&	1.57E+07	&	0.017	\\	
222.5772	&	55418	&	3.5	&	100332	&	4.5	&	0.42	&	3.50E+09	&	0.950	\\	
222.6867	&	54263	&	3.5	&	99155	&	4.5	&	-0.27	&	7.25E+08	&	0.183	\\	
222.7183	&	54557	&	4.5	&	99443	&	4.5	&	0.32	&	2.84E+09	&	0.769	\\	
222.7672	&	56371	&	3.5	&	101247	&	2.5	&	-0.43	&	4.97E+08	&	0.858	\\	
222.8241	&	54176	&	0.5	&	99041	&	1.5	&	-2.29	&	6.93E+06	&	0.018	\\	
222.9781	&	56424	&	1.5	&	101258	&	1.5	&	0.06	&	1.56E+09	&	0.779	\\	
223.0310	&	56424	&	1.5	&	101247	&	2.5	&	-0.27	&	7.28E+08	&	0.429	\\	
223.0370	&	56424	&	1.5	&	101246	&	0.5	&	-1.69	&	2.75E+07	&	0.297	\\	
223.2272	&	54557	&	4.5	&	99341	&	3.5	&	-1.87	&	1.81E+07	&	0.033	\\	
223.2784	&	56371	&	3.5	&	101145	&	3.5	&	-1.31	&	6.48E+07	&	0.073	\\	
223.8459	&	57081	&	3.5	&	101740	&	3.5	&	-0.43	&	4.96E+08	&	0.235	\\	
224.1569	&	54557	&	4.5	&	99155	&	4.5	&	-0.03	&	1.25E+09	&	0.348	\\	
224.2677	&	54557	&	4.5	&	99133	&	5.5	&	0.69	&	6.47E+09	&	0.972	\\	
224.4445	&	55019	&	2.5	&	99559	&	2.5	&	-3.35	&	5.98E+05	&	0.001	\\	
224.4861	&	58493	&	2.5	&	103026	&	2.5	&	0.09	&	1.63E+09	&	0.685	\\	
224.5600	&	56075	&	2.5	&	100593	&	3.5	&	-0.77	&	2.23E+08	&	0.135	\\	
225.0494	&	56424	&	1.5	&	100845	&	0.5	&	-0.73	&	2.47E+08	&	0.672	\\	
225.0759	&	56075	&	2.5	&	100491	&	1.5	&	-0.68	&	2.74E+08	&	0.408	\\	
225.0926	&	54557	&	4.5	&	98969	&	5.5	&	-2.19	&	8.53E+06	&	0.002	\\	
225.1526	&	56075	&	2.5	&	100476	&	3.5	&	-0.36	&	5.72E+08	&	0.171	\\	
225.5525	&	55019	&	2.5	&	99341	&	3.5	&	-1.81	&	2.05E+07	&	0.028	\\	
225.5613	&	57420	&	2.5	&	101740	&	3.5	&	-0.71	&	2.54E+08	&	0.179	\\	
225.5638	&	58706	&	1.5	&	103026	&	2.5	&	-0.46	&	4.59E+08	&	0.695	\\	
225.5911	&	56075	&	2.5	&	100390	&	2.5	&	-0.06	&	1.14E+09	&	0.414	\\	
225.6709	&	54263	&	3.5	&	98561	&	2.5	&	-1.35	&	5.87E+07	&	0.151	\\	
225.7372	&	57081	&	3.5	&	101366	&	2.5	&	-1.68	&	2.74E+07	&	0.073	\\	
225.7828	&	57081	&	3.5	&	101357	&	4.5	&	0.29	&	2.54E+09	&	0.970	\\	
225.9297	&	56371	&	3.5	&	100619	&	4.5	&	0.12	&	1.71E+09	&	0.545	\\	
226.0640	&	56371	&	3.5	&	100593	&	3.5	&	-0.52	&	3.92E+08	&	0.350	\\	
226.1507	&	54263	&	3.5	&	98467	&	3.5	&	-1.75	&	2.33E+07	&	0.040	\\	
226.3443	&	57081	&	3.5	&	101247	&	2.5	&	-1.10	&	1.03E+08	&	0.168	\\	
226.4653	&	55300	&	4.5	&	99443	&	4.5	&	-3.52	&	3.97E+05	&	0.000	\\	
226.4741	&	55418	&	3.5	&	99559	&	2.5	&	-0.27	&	7.12E+08	&	0.428	\\	
226.6646	&	56371	&	3.5	&	100476	&	3.5	&	-0.63	&	3.06E+08	&	0.274	\\	
226.8598	&	56424	&	1.5	&	100491	&	1.5	&	-2.82	&	1.97E+06	&	0.004	\\	
226.8721	&	57081	&	3.5	&	101145	&	3.5	&	-2.40	&	5.11E+06	&	0.008	\\	
226.9915	&	55300	&	4.5	&	99341	&	3.5	&	-1.97	&	1.40E+07	&	0.017	\\	
227.0733	&	55418	&	3.5	&	99443	&	4.5	&	-0.32	&	6.27E+08	&	0.215	\\	
227.0887	&	55019	&	2.5	&	99041	&	1.5	&	-2.15	&	9.02E+06	&	0.079	\\	
227.1090	&	56371	&	3.5	&	100390	&	2.5	&	-1.29	&	6.68E+07	&	0.141	\\	
227.1843	&	56075	&	2.5	&	100079	&	1.5	&	-0.61	&	3.17E+08	&	0.373	\\	
227.3832	&	56424	&	1.5	&	100390	&	2.5	&	-1.43	&	4.84E+07	&	0.031	\\	
227.4057	&	56371	&	3.5	&	100332	&	4.5	&	-3.93	&	1.52E+05	&	0.000	\\	
227.4818	&	57420	&	2.5	&	101366	&	2.5	&	-0.90	&	1.60E+08	&	0.541	\\	
227.6023	&	55418	&	3.5	&	99341	&	3.5	&	-0.03	&	1.20E+09	&	0.364	\\	
227.6672	&	54557	&	4.5	&	98467	&	3.5	&	0.09	&	1.59E+09	&	0.901	\\	
227.9529	&	55300	&	4.5	&	99155	&	4.5	&	-1.81	&	2.00E+07	&	0.006	\\	
228.0430	&	57420	&	2.5	&	101258	&	1.5	&	-3.58	&	3.37E+05	&	0.021	\\	
228.0675	&	55300	&	4.5	&	99133	&	5.5	&	-1.61	&	3.17E+07	&	0.007	\\	
228.0983	&	57420	&	2.5	&	101247	&	2.5	&	-0.12	&	9.78E+08	&	0.702	\\	
228.5689	&	55418	&	3.5	&	99155	&	4.5	&	-0.51	&	4.01E+08	&	0.079	\\	
228.6343	&	57420	&	2.5	&	101145	&	3.5	&	-2.21	&	7.89E+06	&	0.036	\\	
228.9207	&	55300	&	4.5	&	98969	&	5.5	&	-1.49	&	4.15E+07	&	0.005	\\	
229.0019	&	56424	&	1.5	&	100079	&	1.5	&	-1.66	&	2.80E+07	&	0.059	\\	
229.3625	&	56424	&	1.5	&	100010	&	0.5	&	-0.99	&	1.31E+08	&	0.255	\\	
229.5899	&	55019	&	2.5	&	98561	&	2.5	&	-2.71	&	2.50E+06	&	0.019	\\	
229.6099	&	57081	&	3.5	&	100619	&	4.5	&	0.04	&	1.37E+09	&	0.373	\\	
229.7486	&	57081	&	3.5	&	100593	&	3.5	&	-0.20	&	7.98E+08	&	0.486	\\	
230.0865	&	55019	&	2.5	&	98467	&	3.5	&	-3.33	&	5.96E+05	&	0.008	\\	
230.3689	&	57081	&	3.5	&	100476	&	3.5	&	-0.13	&	9.27E+08	&	0.437	\\	
230.8280	&	57081	&	3.5	&	100390	&	2.5	&	-1.29	&	6.46E+07	&	0.116	\\	
231.0611	&	56075	&	2.5	&	99341	&	3.5	&	-1.66	&	2.73E+07	&	0.015	\\	
231.1346	&	57081	&	3.5	&	100332	&	4.5	&	-0.69	&	2.56E+08	&	0.100	\\	
231.1585	&	58493	&	2.5	&	101740	&	3.5	&	0.37	&	2.88E+09	&	0.937	\\	
231.5560	&	57420	&	2.5	&	100593	&	3.5	&	-0.41	&	4.80E+08	&	0.666	\\	
231.5840	&	55300	&	4.5	&	98467	&	3.5	&	-1.37	&	5.36E+07	&	0.130	\\	
231.7140	&	55418	&	3.5	&	98561	&	2.5	&	-1.01	&	1.22E+08	&	0.136	\\	
231.7599	&	56424	&	1.5	&	99559	&	2.5	&	-2.56	&	3.43E+06	&	0.009	\\	
232.1011	&	56371	&	3.5	&	99443	&	4.5	&	-0.65	&	2.79E+08	&	0.136	\\	
232.1046	&	57420	&	2.5	&	100491	&	1.5	&	-0.82	&	1.86E+08	&	0.575	\\	
232.1862	&	57420	&	2.5	&	100476	&	3.5	&	-0.85	&	1.74E+08	&	0.211	\\	
232.2198	&	55418	&	3.5	&	98467	&	3.5	&	-1.40	&	4.92E+07	&	0.037	\\	
232.6525	&	57420	&	2.5	&	100390	&	2.5	&	-0.39	&	5.00E+08	&	0.647	\\	
232.6537	&	56371	&	3.5	&	99341	&	3.5	&	-0.76	&	2.13E+08	&	0.156	\\	
232.6735	&	56075	&	2.5	&	99041	&	1.5	&	-1.54	&	3.57E+07	&	0.073	\\	
233.1759	&	58493	&	2.5	&	101366	&	2.5	&	-0.96	&	1.34E+08	&	0.207	\\	
233.6639	&	56371	&	3.5	&	99155	&	4.5	&	-1.76	&	2.13E+07	&	0.022	\\	
233.7655	&	58493	&	2.5	&	101258	&	1.5	&	-3.41	&	4.75E+05	&	0.008	\\	
233.8237	&	58493	&	2.5	&	101247	&	2.5	&	-0.35	&	5.39E+08	&	0.250	\\	
234.3388	&	58706	&	1.5	&	101366	&	2.5	&	-0.84	&	1.75E+08	&	0.533	\\	
234.3474	&	57420	&	2.5	&	100079	&	1.5	&	-0.11	&	9.41E+08	&	0.948	\\	
234.3870	&	58493	&	2.5	&	101145	&	3.5	&	-1.85	&	1.69E+07	&	0.026	\\	
234.5804	&	56424	&	1.5	&	99041	&	1.5	&	-2.44	&	4.44E+06	&	0.019	\\	
234.9343	&	58706	&	1.5	&	101258	&	1.5	&	-1.58	&	3.16E+07	&	0.335	\\	
234.9931	&	58706	&	1.5	&	101247	&	2.5	&	-0.64	&	2.73E+08	&	0.259	\\	
234.9998	&	58706	&	1.5	&	101246	&	0.5	&	-0.16	&	8.32E+08	&	0.917	\\	
235.2999	&	56075	&	2.5	&	98561	&	2.5	&	-2.04	&	1.11E+07	&	0.022	\\	
235.3396	&	57081	&	3.5	&	99559	&	2.5	&	-2.27	&	6.45E+06	&	0.015	\\	
235.8215	&	56075	&	2.5	&	98467	&	3.5	&	-2.74	&	2.19E+06	&	0.008	\\	
235.9867	&	57081	&	3.5	&	99443	&	4.5	&	-0.18	&	7.82E+08	&	0.259	\\	
236.5581	&	57081	&	3.5	&	99341	&	3.5	&	-0.15	&	8.49E+08	&	0.251	\\	
236.9517	&	56371	&	3.5	&	98561	&	2.5	&	-2.31	&	5.79E+06	&	0.461	\\	
237.2348	&	58706	&	1.5	&	100845	&	0.5	&	-0.90	&	1.49E+08	&	0.420	\\	
237.2365	&	57420	&	2.5	&	99559	&	2.5	&	-1.38	&	4.95E+07	&	0.130	\\	
237.2502	&	56424	&	1.5	&	98561	&	2.5	&	-3.15	&	8.43E+05	&	0.010	\\	
237.4585	&	58493	&	2.5	&	100593	&	3.5	&	-1.14	&	8.48E+07	&	0.073	\\	
237.4807	&	56371	&	3.5	&	98467	&	3.5	&	-2.16	&	8.18E+06	&	0.168	\\	
237.6025	&	57081	&	3.5	&	99155	&	4.5	&	-1.24	&	6.81E+07	&	0.037	\\	
238.0354	&	58493	&	2.5	&	100491	&	1.5	&	-1.42	&	4.45E+07	&	0.214	\\	
238.1212	&	58493	&	2.5	&	100476	&	3.5	&	-0.66	&	2.55E+08	&	0.256	\\	
238.4747	&	57420	&	2.5	&	99341	&	3.5	&	-0.26	&	6.41E+08	&	0.430	\\	
238.6117	&	58493	&	2.5	&	100390	&	2.5	&	-0.80	&	1.85E+08	&	0.118	\\	
239.2474	&	58706	&	1.5	&	100491	&	1.5	&	-2.89	&	1.51E+06	&	0.008	\\	
239.8296	&	58706	&	1.5	&	100390	&	2.5	&	-0.57	&	3.08E+08	&	0.426	\\	
240.1926	&	57420	&	2.5	&	99041	&	1.5	&	-0.12	&	8.73E+08	&	0.653	\\	
240.3949	&	58493	&	2.5	&	100079	&	1.5	&	-0.95	&	1.29E+08	&	0.288	\\	
241.0028	&	57081	&	3.5	&	98561	&	2.5	&	-1.63	&	2.70E+07	&	0.268	\\	
241.5501	&	57081	&	3.5	&	98467	&	3.5	&	-1.63	&	2.68E+07	&	0.100	\\	
241.6311	&	58706	&	1.5	&	100079	&	1.5	&	-1.23	&	6.68E+07	&	0.297	\\	
242.0325	&	58706	&	1.5	&	100010	&	0.5	&	-0.67	&	2.44E+08	&	0.626	\\	
242.9924	&	57420	&	2.5	&	98561	&	2.5	&	-3.12	&	8.66E+05	&	0.021	\\	
243.4359	&	58493	&	2.5	&	99559	&	2.5	&	-3.33	&	5.33E+05	&	0.001	\\	
243.5488	&	57420	&	2.5	&	98467	&	3.5	&	-2.29	&	5.78E+06	&	0.097	\\	
244.7036	&	58706	&	1.5	&	99559	&	2.5	&	-2.07	&	9.57E+06	&	0.071	\\	
244.7398	&	58493	&	2.5	&	99341	&	3.5	&	-1.07	&	9.53E+07	&	0.063	\\	
246.5495	&	58493	&	2.5	&	99041	&	1.5	&	-0.66	&	2.42E+08	&	0.418	\\	
247.8500	&	58706	&	1.5	&	99041	&	1.5	&	-1.33	&	5.07E+07	&	0.147	\\	
250.0870	&	58493	&	2.5	&	98467	&	3.5	&	-2.75	&	1.91E+06	&	0.036	\\	
269.5188	&	66571	&	2.5	&	103664	&	1.5	&	-3.65	&	2.05E+05	&	0.005	\\	
269.5796	&	66580	&	1.5	&	103664	&	1.5	&	-2.51	&	2.83E+06	&	0.025	\\	
272.9010	&	67031	&	0.5	&	103664	&	1.5	&	-2.86	&	1.25E+06	&	0.077	\\	
274.2354	&	66571	&	2.5	&	103026	&	2.5	&	-2.27	&	4.75E+06	&	0.040	\\	
274.2984	&	66580	&	1.5	&	103026	&	2.5	&	-2.13	&	6.59E+06	&	0.120	\\	
277.9362	&	67695	&	2.5	&	103664	&	1.5	&	-2.19	&	5.63E+06	&	0.080	\\	
281.5343	&	68154	&	1.5	&	103664	&	1.5	&	-1.64	&	1.91E+07	&	0.060	\\	
282.5474	&	68282	&	0.5	&	103664	&	1.5	&	-2.28	&	4.42E+06	&	0.081	\\	
282.9548	&	67695	&	2.5	&	103026	&	2.5	&	-2.48	&	2.76E+06	&	0.055	\\	
286.2231	&	68736	&	2.5	&	103664	&	1.5	&	-3.57	&	2.19E+05	&	0.003	\\	
286.4951	&	68131	&	3.5	&	103026	&	2.5	&	-1.94	&	9.48E+06	&	0.227	\\	
286.6848	&	68154	&	1.5	&	103026	&	2.5	&	-2.04	&	7.41E+06	&	0.081	\\	
287.3149	&	66571	&	2.5	&	101366	&	2.5	&	-3.36	&	3.55E+05	&	0.013	\\	
288.1178	&	68966	&	1.5	&	103664	&	1.5	&	-1.90	&	1.01E+07	&	0.096	\\	
288.2991	&	66571	&	2.5	&	101247	&	2.5	&	-2.44	&	2.93E+06	&	0.074	\\	
288.3687	&	66580	&	1.5	&	101247	&	2.5	&	-2.60	&	2.00E+06	&	0.056	\\	
288.3788	&	66580	&	1.5	&	101246	&	0.5	&	-2.10	&	6.30E+06	&	0.250	\\	
289.1558	&	66571	&	2.5	&	101145	&	3.5	&	-3.68	&	1.66E+05	&	0.012	\\	
291.5482	&	68736	&	2.5	&	103026	&	2.5	&	-2.28	&	4.08E+06	&	0.020	\\	
291.7516	&	66580	&	1.5	&	100845	&	0.5	&	-2.37	&	3.32E+06	&	0.213	\\	
292.0816	&	67031	&	0.5	&	101258	&	1.5	&	-3.25	&	4.43E+05	&	0.145	\\	
292.1827	&	67031	&	0.5	&	101246	&	0.5	&	-3.77	&	1.31E+05	&	0.070	\\	
293.5143	&	68966	&	1.5	&	103026	&	2.5	&	-2.26	&	4.27E+06	&	0.062	\\	
293.6376	&	67695	&	2.5	&	101740	&	3.5	&	-2.04	&	7.09E+06	&	0.183	\\	
293.8446	&	66571	&	2.5	&	100593	&	3.5	&	-4.00	&	7.73E+04	&	0.003	\\	
294.8601	&	66571	&	2.5	&	100476	&	3.5	&	-2.51	&	2.34E+06	&	0.107	\\	
295.6126	&	66571	&	2.5	&	100390	&	2.5	&	-2.82	&	1.15E+06	&	0.049	\\	
295.6457	&	67031	&	0.5	&	100845	&	0.5	&	-3.92	&	9.29E+04	&	0.022	\\	
295.6857	&	66580	&	1.5	&	100390	&	2.5	&	-2.08	&	6.30E+06	&	0.330	\\	
296.9004	&	67695	&	2.5	&	101366	&	2.5	&	-2.82	&	1.16E+06	&	0.038	\\	
297.4520	&	68131	&	3.5	&	101740	&	3.5	&	-3.69	&	1.52E+05	&	0.004	\\	
297.9514	&	67695	&	2.5	&	101247	&	2.5	&	-2.12	&	5.68E+06	&	0.128	\\	
298.3541	&	66571	&	2.5	&	100079	&	1.5	&	-1.98	&	7.77E+06	&	0.297	\\	
298.7778	&	67031	&	0.5	&	100491	&	1.5	&	-3.04	&	6.86E+05	&	0.156	\\	
298.8666	&	67695	&	2.5	&	101145	&	3.5	&	-3.25	&	4.17E+05	&	0.006	\\	
299.0412	&	66580	&	1.5	&	100010	&	0.5	&	-2.48	&	2.46E+06	&	0.213	\\	
300.8815	&	68131	&	3.5	&	101357	&	4.5	&	-3.10	&	5.90E+05	&	0.007	\\	
301.0098	&	68154	&	1.5	&	101366	&	2.5	&	-2.66	&	1.61E+06	&	0.035	\\	
302.0902	&	68154	&	1.5	&	101247	&	2.5	&	-1.72	&	1.39E+07	&	0.148	\\	
302.1012	&	68154	&	1.5	&	101246	&	0.5	&	-1.44	&	2.63E+07	&	0.534	\\	
302.5043	&	67031	&	0.5	&	100079	&	1.5	&	-3.19	&	4.67E+05	&	0.191	\\	
302.6849	&	70635	&	2.5	&	103664	&	1.5	&	-3.16	&	5.07E+05	&	0.042	\\	
302.9027	&	68736	&	2.5	&	101740	&	3.5	&	-2.19	&	4.67E+06	&	0.096	\\	
303.0524	&	66571	&	2.5	&	99559	&	2.5	&	-2.74	&	1.34E+06	&	0.037	\\	
303.1293	&	66580	&	1.5	&	99559	&	2.5	&	-2.70	&	1.47E+06	&	0.093	\\	
303.1336	&	67031	&	0.5	&	100010	&	0.5	&	-3.14	&	5.32E+05	&	0.187	\\	
303.3399	&	70707	&	1.5	&	103664	&	1.5	&	-3.22	&	4.44E+05	&	0.038	\\	
303.8783	&	67695	&	2.5	&	100593	&	3.5	&	-2.76	&	1.27E+06	&	0.033	\\	
304.8237	&	67695	&	2.5	&	100491	&	1.5	&	-3.68	&	1.50E+05	&	0.016	\\	
305.0758	&	66571	&	2.5	&	99341	&	3.5	&	-1.87	&	9.75E+06	&	0.338	\\	
305.7694	&	67695	&	2.5	&	100390	&	2.5	&	-2.60	&	1.80E+06	&	0.077	\\	
305.8047	&	68154	&	1.5	&	100845	&	0.5	&	-1.99	&	7.23E+06	&	0.268	\\	
306.3759	&	68736	&	2.5	&	101366	&	2.5	&	-3.27	&	3.79E+05	&	0.010	\\	
307.0003	&	68282	&	0.5	&	100845	&	0.5	&	-3.54	&	2.04E+05	&	0.024	\\	
307.4952	&	68736	&	2.5	&	101247	&	2.5	&	-2.67	&	1.50E+06	&	0.026	\\	
307.7161	&	68131	&	3.5	&	100619	&	4.5	&	-2.26	&	3.91E+06	&	0.132	\\	
307.8927	&	66571	&	2.5	&	99041	&	1.5	&	-1.66	&	1.54E+07	&	0.486	\\	
307.9652	&	68131	&	3.5	&	100593	&	3.5	&	-2.60	&	1.75E+06	&	0.078	\\	
307.9721	&	66580	&	1.5	&	99041	&	1.5	&	-2.65	&	1.56E+06	&	0.120	\\	
308.4700	&	68736	&	2.5	&	101145	&	3.5	&	-3.41	&	2.68E+05	&	0.014	\\	
308.5477	&	68966	&	1.5	&	101366	&	2.5	&	-3.45	&	2.50E+05	&	0.009	\\	
308.7035	&	67695	&	2.5	&	100079	&	1.5	&	-3.91	&	8.49E+04	&	0.012	\\	
309.0808	&	68131	&	3.5	&	100476	&	3.5	&	-2.48	&	2.29E+06	&	0.165	\\	
309.1570	&	68154	&	1.5	&	100491	&	1.5	&	-3.82	&	1.06E+05	&	0.006	\\	
309.6829	&	68966	&	1.5	&	101247	&	2.5	&	-2.29	&	3.59E+06	&	0.119	\\	
309.6946	&	68966	&	1.5	&	101246	&	0.5	&	-2.35	&	3.07E+06	&	0.202	\\	
309.9077	&	68131	&	3.5	&	100390	&	2.5	&	-3.71	&	1.35E+05	&	0.015	\\	
310.0120	&	70778	&	3.5	&	103026	&	2.5	&	-2.73	&	1.30E+06	&	0.089	\\	
310.1298	&	68154	&	1.5	&	100390	&	2.5	&	-1.88	&	9.15E+06	&	0.207	\\	
310.4604	&	68131	&	3.5	&	100332	&	4.5	&	-2.85	&	9.72E+05	&	0.037	\\	
312.3145	&	67031	&	0.5	&	99041	&	1.5	&	-3.29	&	3.48E+05	&	0.161	\\	
312.5083	&	66571	&	2.5	&	98561	&	2.5	&	-3.40	&	2.75E+05	&	0.020	\\	
312.5900	&	66580	&	1.5	&	98561	&	2.5	&	-2.92	&	8.34E+05	&	0.183	\\	
313.1486	&	68154	&	1.5	&	100079	&	1.5	&	-2.16	&	4.69E+06	&	0.273	\\	
313.4290	&	66571	&	2.5	&	98467	&	3.5	&	-2.02	&	6.46E+06	&	0.433	\\	
313.4608	&	71771	&	2.5	&	103664	&	1.5	&	-2.58	&	1.77E+06	&	0.171	\\	
313.5878	&	68966	&	1.5	&	100845	&	0.5	&	-2.75	&	1.21E+06	&	0.080	\\	
313.7361	&	67695	&	2.5	&	99559	&	2.5	&	-2.29	&	3.50E+06	&	0.249	\\	
313.8118	&	68736	&	2.5	&	100593	&	3.5	&	-2.45	&	2.38E+06	&	0.129	\\	
313.8231	&	68154	&	1.5	&	100010	&	0.5	&	-1.85	&	9.65E+06	&	0.415	\\	
314.4024	&	68282	&	0.5	&	100079	&	1.5	&	-3.52	&	2.05E+05	&	0.065	\\	
314.8202	&	68736	&	2.5	&	100491	&	1.5	&	-2.42	&	2.53E+06	&	0.148	\\	
314.9702	&	68736	&	2.5	&	100476	&	3.5	&	-1.80	&	1.06E+07	&	0.355	\\	
315.0823	&	68282	&	0.5	&	100010	&	0.5	&	-3.66	&	1.48E+05	&	0.035	\\	
315.8290	&	68736	&	2.5	&	100390	&	2.5	&	-3.52	&	1.99E+05	&	0.011	\\	
315.9051	&	67695	&	2.5	&	99341	&	3.5	&	-2.79	&	1.08E+06	&	0.060	\\	
318.0944	&	68131	&	3.5	&	99559	&	2.5	&	-2.48	&	2.22E+06	&	0.099	\\	
318.1374	&	68966	&	1.5	&	100390	&	2.5	&	-2.43	&	2.47E+06	&	0.220	\\	
318.3284	&	68154	&	1.5	&	99559	&	2.5	&	-2.88	&	8.74E+05	&	0.055	\\	
318.9266	&	67695	&	2.5	&	99041	&	1.5	&	-2.35	&	2.93E+06	&	0.290	\\	
318.9603	&	68736	&	2.5	&	100079	&	1.5	&	-1.54	&	1.85E+07	&	0.617	\\	
319.2776	&	68131	&	3.5	&	99443	&	4.5	&	-2.77	&	1.11E+06	&	0.032	\\	
319.5182	&	72375	&	1.5	&	103664	&	1.5	&	-1.67	&	1.40E+07	&	0.533	\\	
319.8589	&	71771	&	2.5	&	103026	&	2.5	&	-1.51	&	2.02E+07	&	0.581	\\	
320.3243	&	68131	&	3.5	&	99341	&	3.5	&	-2.37	&	2.78E+06	&	0.183	\\	
321.3149	&	68966	&	1.5	&	100079	&	1.5	&	-3.15	&	4.62E+05	&	0.056	\\	
321.4008	&	70635	&	2.5	&	101740	&	3.5	&	-3.74	&	1.18E+05	&	0.006	\\	
322.0250	&	68966	&	1.5	&	100010	&	0.5	&	-2.78	&	1.07E+06	&	0.128	\\	
322.2422	&	68131	&	3.5	&	99155	&	4.5	&	-2.57	&	1.73E+06	&	0.077	\\	
323.6733	&	68154	&	1.5	&	99041	&	1.5	&	-2.61	&	1.55E+06	&	0.066	\\	
324.3358	&	68736	&	2.5	&	99559	&	2.5	&	-3.02	&	6.05E+05	&	0.024	\\	
325.8741	&	72986	&	1.5	&	103664	&	1.5	&	-2.84	&	9.27E+05	&	0.028	\\	
326.0705	&	70707	&	1.5	&	101366	&	2.5	&	-2.17	&	4.25E+06	&	0.129	\\	
326.1685	&	72375	&	1.5	&	103026	&	2.5	&	-2.82	&	9.55E+05	&	0.062	\\	
326.5761	&	70635	&	2.5	&	101247	&	2.5	&	-3.52	&	1.88E+05	&	0.019	\\	
326.6544	&	68736	&	2.5	&	99341	&	3.5	&	-1.67	&	1.33E+07	&	0.399	\\	
327.2246	&	70707	&	1.5	&	101258	&	1.5	&	-3.11	&	4.83E+05	&	0.021	\\	
327.3386	&	70707	&	1.5	&	101247	&	2.5	&	-2.70	&	1.25E+06	&	0.154	\\	
327.3516	&	70707	&	1.5	&	101246	&	0.5	&	-1.87	&	8.33E+06	&	0.764	\\	
327.6744	&	70749	&	0.5	&	101258	&	1.5	&	-2.12	&	4.76E+06	&	0.154	\\	
327.6759	&	70635	&	2.5	&	101145	&	3.5	&	-2.95	&	6.99E+05	&	0.056	\\	
327.8017	&	70749	&	0.5	&	101246	&	0.5	&	-2.31	&	3.06E+06	&	0.407	\\	
328.5283	&	68131	&	3.5	&	98561	&	2.5	&	-1.87	&	8.37E+06	&	0.517	\\	
328.7779	&	68154	&	1.5	&	98561	&	2.5	&	-3.87	&	8.27E+04	&	0.036	\\	
329.5460	&	68131	&	3.5	&	98467	&	3.5	&	-2.14	&	4.45E+06	&	0.249	\\	
329.8861	&	68736	&	2.5	&	99041	&	1.5	&	-1.44	&	2.22E+07	&	0.605	\\	
331.7045	&	70707	&	1.5	&	100845	&	0.5	&	-3.07	&	5.14E+05	&	0.022	\\	
332.1667	&	70749	&	0.5	&	100845	&	0.5	&	-1.62	&	1.47E+07	&	0.721	\\	
332.4053	&	68966	&	1.5	&	99041	&	1.5	&	-2.98	&	6.32E+05	&	0.060	\\	
332.7945	&	72986	&	1.5	&	103026	&	2.5	&	-1.26	&	3.38E+07	&	0.710	\\	
333.5773	&	71771	&	2.5	&	101740	&	3.5	&	-3.50	&	1.89E+05	&	0.029	\\	
333.7100	&	70635	&	2.5	&	100593	&	3.5	&	-2.05	&	5.33E+06	&	0.124	\\	
334.8505	&	70635	&	2.5	&	100491	&	1.5	&	-1.79	&	9.63E+06	&	0.233	\\	
335.0203	&	70635	&	2.5	&	100476	&	3.5	&	-3.09	&	4.86E+05	&	0.077	\\	
335.1901	&	68736	&	2.5	&	98561	&	2.5	&	-3.67	&	1.28E+05	&	0.028	\\	
335.3068	&	70778	&	3.5	&	100593	&	3.5	&	-3.34	&	2.75E+05	&	0.013	\\	
335.6522	&	70707	&	1.5	&	100491	&	1.5	&	-1.45	&	2.12E+07	&	0.583	\\	
335.9221	&	73903	&	0.5	&	103664	&	1.5	&	-1.57	&	1.60E+07	&	0.586	\\	
335.9920	&	70635	&	2.5	&	100390	&	2.5	&	-2.15	&	4.22E+06	&	0.178	\\	
336.1255	&	70749	&	0.5	&	100491	&	1.5	&	-2.20	&	3.71E+06	&	0.248	\\	
336.7992	&	70707	&	1.5	&	100390	&	2.5	&	-2.51	&	1.84E+06	&	0.084	\\	
337.6108	&	70778	&	3.5	&	100390	&	2.5	&	-3.69	&	1.19E+05	&	0.013	\\	
338.2669	&	70778	&	3.5	&	100332	&	4.5	&	-3.20	&	3.67E+05	&	0.096	\\	
339.1555	&	71771	&	2.5	&	101247	&	2.5	&	-3.41	&	2.23E+05	&	0.033	\\	
339.5382	&	70635	&	2.5	&	100079	&	1.5	&	-1.75	&	1.02E+07	&	0.573	\\	
340.2680	&	74283	&	0.5	&	103664	&	1.5	&	-2.50	&	1.80E+06	&	0.445	\\	
340.3625	&	70707	&	1.5	&	100079	&	1.5	&	-2.75	&	1.03E+06	&	0.053	\\	
340.4719	&	74301	&	1.5	&	103664	&	1.5	&	-3.68	&	1.23E+05	&	0.055	\\	
340.8491	&	70749	&	0.5	&	100079	&	1.5	&	-2.86	&	7.98E+05	&	0.068	\\	
341.1595	&	70707	&	1.5	&	100010	&	0.5	&	-1.88	&	7.59E+06	&	0.440	\\	
341.6483	&	70749	&	0.5	&	100010	&	0.5	&	-3.51	&	1.77E+05	&	0.011	\\	
345.6362	&	70635	&	2.5	&	99559	&	2.5	&	-1.45	&	2.02E+07	&	0.414	\\	
346.2577	&	72375	&	1.5	&	101247	&	2.5	&	-3.43	&	2.03E+05	&	0.027	\\	
346.4905	&	70707	&	1.5	&	99559	&	2.5	&	-2.25	&	3.17E+06	&	0.115	\\	
347.3494	&	70778	&	3.5	&	99559	&	2.5	&	-2.02	&	5.26E+06	&	0.125	\\	
348.0334	&	74301	&	1.5	&	103026	&	2.5	&	-2.64	&	1.27E+06	&	0.415	\\	
348.0883	&	71771	&	2.5	&	100491	&	1.5	&	-2.58	&	1.44E+06	&	0.209	\\	
348.2706	&	70635	&	2.5	&	99341	&	3.5	&	-3.72	&	1.06E+05	&	0.006	\\	
348.7608	&	70778	&	3.5	&	99443	&	4.5	&	-2.13	&	4.08E+06	&	0.122	\\	
349.3220	&	71771	&	2.5	&	100390	&	2.5	&	-3.39	&	2.21E+05	&	0.038	\\	
350.0101	&	70778	&	3.5	&	99341	&	3.5	&	-2.75	&	9.61E+05	&	0.116	\\	
351.9465	&	70635	&	2.5	&	99041	&	1.5	&	-1.85	&	7.70E+06	&	0.373	\\	
352.2540	&	72986	&	1.5	&	101366	&	2.5	&	-3.75	&	9.65E+04	&	0.020	\\	
352.3012	&	70778	&	3.5	&	99155	&	4.5	&	-2.98	&	5.71E+05	&	0.021	\\	
353.1567	&	71771	&	2.5	&	100079	&	1.5	&	-2.45	&	1.87E+06	&	0.260	\\	
353.3552	&	70749	&	0.5	&	99041	&	1.5	&	-3.98	&	5.59E+04	&	0.009	\\	
353.7344	&	72986	&	1.5	&	101247	&	2.5	&	-2.88	&	7.13E+05	&	0.079	\\	
353.7495	&	72986	&	1.5	&	101246	&	0.5	&	-3.08	&	4.44E+05	&	0.073	\\	
355.5737	&	72375	&	1.5	&	100491	&	1.5	&	-3.92	&	6.29E+04	&	0.026	\\	
356.8612	&	72375	&	1.5	&	100390	&	2.5	&	-2.97	&	5.60E+05	&	0.143	\\	
358.9066	&	70707	&	1.5	&	98561	&	2.5	&	-3.61	&	1.28E+05	&	0.010	\\	
359.1988	&	70635	&	2.5	&	98467	&	3.5	&	-2.61	&	1.29E+06	&	0.067	\\	
359.7584	&	71771	&	2.5	&	99559	&	2.5	&	-2.97	&	5.57E+05	&	0.065	\\	
359.8284	&	70778	&	3.5	&	98561	&	2.5	&	-1.24	&	3.01E+07	&	0.685	\\	
361.0496	&	70778	&	3.5	&	98467	&	3.5	&	-1.40	&	2.07E+07	&	0.446	\\	
361.7601	&	72375	&	1.5	&	100010	&	0.5	&	-3.40	&	2.04E+05	&	0.149	\\	
362.6134	&	71771	&	2.5	&	99341	&	3.5	&	-3.35	&	2.26E+05	&	0.022	\\	
363.4627	&	72986	&	1.5	&	100491	&	1.5	&	-3.73	&	9.59E+04	&	0.019	\\	
364.8080	&	72986	&	1.5	&	100390	&	2.5	&	-3.71	&	9.78E+04	&	0.026	\\	
366.6000	&	71771	&	2.5	&	99041	&	1.5	&	-3.99	&	5.08E+04	&	0.006	\\	
366.7136	&	76402	&	2.5	&	103664	&	1.5	&	-2.65	&	1.12E+06	&	0.020	\\	
368.7905	&	75918	&	3.5	&	103026	&	2.5	&	-2.33	&	2.30E+06	&	0.037	\\	
370.7754	&	74283	&	0.5	&	101246	&	0.5	&	-3.34	&	2.16E+05	&	0.216	\\	
374.4756	&	71771	&	2.5	&	98467	&	3.5	&	-3.45	&	1.68E+05	&	0.065	\\	
374.9122	&	72375	&	1.5	&	99041	&	1.5	&	-3.79	&	7.70E+04	&	0.041	\\	
375.5006	&	76402	&	2.5	&	103026	&	2.5	&	-3.01	&	4.68E+05	&	0.023	\\	
376.0069	&	73903	&	0.5	&	100491	&	1.5	&	-3.45	&	1.68E+05	&	0.101	\\	
376.2053	&	72986	&	1.5	&	99559	&	2.5	&	-3.12	&	3.60E+05	&	0.080	\\	
376.3696	&	74283	&	0.5	&	100845	&	0.5	&	-3.97	&	4.95E+04	&	0.062	\\	
381.4587	&	75149	&	4.5	&	101357	&	4.5	&	-3.79	&	7.50E+04	&	0.100	\\	
381.9278	&	73903	&	0.5	&	100079	&	1.5	&	-3.81	&	7.14E+04	&	0.072	\\	
387.1473	&	75918	&	3.5	&	101740	&	3.5	&	-3.38	&	1.82E+05	&	0.025	\\	
387.5554	&	74283	&	0.5	&	100079	&	1.5	&	-3.49	&	1.41E+05	&	0.197	\\	
388.5890	&	74283	&	0.5	&	100010	&	0.5	&	-3.86	&	6.01E+04	&	0.157	\\	
392.9771	&	75918	&	3.5	&	101357	&	4.5	&	-2.62	&	1.04E+06	&	0.151	\\	
394.5487	&	76402	&	2.5	&	101740	&	3.5	&	-1.57	&	1.15E+07	&	0.545	\\	
394.6811	&	75918	&	3.5	&	101247	&	2.5	&	-3.49	&	1.38E+05	&	0.026	\\	
397.6995	&	73903	&	0.5	&	99041	&	1.5	&	-3.29	&	2.20E+05	&	0.182	\\	
400.4618	&	76402	&	2.5	&	101366	&	2.5	&	-2.81	&	6.39E+05	&	0.234	\\	
402.3762	&	76402	&	2.5	&	101247	&	2.5	&	-1.75	&	7.35E+06	&	0.543	\\	
403.8052	&	74283	&	0.5	&	99041	&	1.5	&	-3.57	&	1.06E+05	&	0.131	\\	
404.0471	&	76402	&	2.5	&	101145	&	3.5	&	-2.48	&	1.34E+06	&	0.294	\\	
404.7172	&	75918	&	3.5	&	100619	&	4.5	&	-1.95	&	4.53E+06	&	0.474	\\	
405.1483	&	75918	&	3.5	&	100593	&	3.5	&	-2.36	&	1.76E+06	&	0.357	\\	
407.0812	&	75918	&	3.5	&	100476	&	3.5	&	-1.95	&	4.51E+06	&	0.596	\\	
408.5168	&	75918	&	3.5	&	100390	&	2.5	&	-3.50	&	1.28E+05	&	0.060	\\	
409.4777	&	75918	&	3.5	&	100332	&	4.5	&	-2.17	&	2.66E+06	&	0.462	\\	
413.2612	&	76402	&	2.5	&	100593	&	3.5	&	-2.35	&	1.75E+06	&	0.229	\\	
415.2724	&	76402	&	2.5	&	100476	&	3.5	&	-2.88	&	5.04E+05	&	0.168	\\	
416.7665	&	76402	&	2.5	&	100390	&	2.5	&	-1.92	&	4.61E+06	&	0.559	\\	
424.9559	&	75918	&	3.5	&	99443	&	4.5	&	-1.83	&	5.43E+06	&	0.394	\\	
426.8121	&	75918	&	3.5	&	99341	&	3.5	&	-1.65	&	8.29E+06	&	0.667	\\	
430.0280	&	75722	&	5.5	&	98969	&	5.5	&	-3.96	&	4.00E+04	&	0.044	\\	
430.2238	&	75918	&	3.5	&	99155	&	4.5	&	-2.46	&	1.26E+06	&	0.237	\\	
431.7079	&	76402	&	2.5	&	99559	&	2.5	&	-3.01	&	3.55E+05	&	0.145	\\	
435.8254	&	76402	&	2.5	&	99341	&	3.5	&	-3.23	&	2.07E+05	&	0.030	\\	
443.3420	&	75918	&	3.5	&	98467	&	3.5	&	-3.84	&	4.90E+04	&	0.015	\\	
456.1340	&	79823	&	3.5	&	101740	&	3.5	&	-3.02	&	3.04E+05	&	0.299	\\	
468.8767	&	79823	&	3.5	&	101145	&	3.5	&	-3.84	&	4.35E+04	&	0.185	\\	
481.3303	&	79823	&	3.5	&	100593	&	3.5	&	-3.51	&	8.85E+04	&	0.188	\\	
483.0646	&	79924	&	4.5	&	100619	&	4.5	&	-3.34	&	1.32E+05	&	0.165	\\	
489.8621	&	79924	&	4.5	&	100332	&	4.5	&	-3.59	&	7.05E+04	&	0.263	\\	
512.1791	&	79924	&	4.5	&	99443	&	4.5	&	-3.30	&	1.29E+05	&	0.095	\\	
512.2174	&	79823	&	3.5	&	99341	&	3.5	&	-3.47	&	8.73E+04	&	0.062	\\	
519.8509	&	79924	&	4.5	&	99155	&	4.5	&	-3.67	&	5.31E+04	&	0.038	\\	
\end{longtable}
\noindent
$^a$ Vacuum ($\lambda$ $<$ 200 nm) and air ($\lambda$ $>$ 200 nm) Ritz wavelengths deduced from the experimental energy level values by \citet{S70}\\
$^b$ Energies from \citet{S70}. Energy values have been rounded to the nearest unit. \\
$^c$ This work.
\end{document}